\documentclass[aps,prd,preprintnumbers,superscriptaddress,nofootinbib,showpacs]{revtex4-1}%

\usepackage[pdftex]{graphicx} 
\usepackage[pdftex]{color}
\input{colordvi.tex}
\usepackage{bm,amsmath,amssymb,amsfonts,mathrsfs}
\usepackage[pdftex]{hyperref}
\usepackage{url}
\hypersetup{
    colorlinks=true,
    citecolor=cyan,
}

\newcommand{\henbi}{\partial}

\begin{document}

\title{Anti-screening of the Galileon force around a disk center hole}
\author{Hiromu~Ogawa}
\email[Email: ]{jh.ogawa"at"rikkyo.ac.jp}
\affiliation{Department of Physics, Rikkyo University, Toshima, Tokyo 171-8501, Japan
}
\affiliation{Institute of Cosmology and Gravitation, University of Portsmouth, Portsmouth, PO1 3FX, UK}

\author{Takashi~Hiramatsu}
\email[Email: ]{hiramatz"at"rikkyo.ac.jp}
\affiliation{Department of Physics, Rikkyo University, Toshima, Tokyo 171-8501, Japan
}

\author{Tsutomu~Kobayashi}
\email[Email: ]{tsutomu"at"rikkyo.ac.jp}
\affiliation{Department of Physics, Rikkyo University, Toshima, Tokyo 171-8501, Japan
}
\begin{abstract}
The Vainshtein mechanism is known as an efficient way of screening the
fifth force around a matter source in modified gravity.
This has been verified mainly in highly symmetric matter configurations.
To study how the Vainshtein mechanism works in a less symmetric setup,
we numerically solve the scalar field equation around
a disk with a hole at its center in the cubic Galileon theory.
We find, surprisingly, that the Galileon force is enhanced,
rather than suppressed, in the vicinity of the hole.
This anti-screening effect is larger for a thinner, less massive disk
with a smaller hole. At this stage our setup is only of academic interest
and its astrophysical consequences are unclear, but this result
implies that the Vainshtein screening mechanism around
less symmetric matter configurations is quite nontrivial.
\end{abstract}
\preprint{RUP-18-5}
\maketitle

\section{Introduction}

The origin of the current accelerated expansion of the
Universe~\cite{Perlmutter:1998np,Riess:1998cb}
is one of the biggest problem in modern physics
(see~\cite{Spergel:2003cb,Eisenstein:2005su,Abazjian:2008wr,Ade:2013zuv} for several
independent observations).
In order to explain this phenomenon, one basically has
to introduce an unknown energy source, i.e., dark energy~\cite{Copeland:2006wr}.
The simplest possibility is the cosmological constant,
and introducing another unknown component, i.e., dark matter,
the $\Lambda$CDM model is well compatible with observations.
This concordance model, however, suffers from the fine-tuning problem
of the cosmological constant.
The mystery of the accelerated expansion of the Universe thus
motivates us to explore the possibilities of
modification of general relativity on cosmological scales~\cite{clifton}.

A number of theories have been proposed so far as alternatives to
the cosmological constant and dark energy.
Most of them can be described (effectively) by a scalar-tensor theory,
which has a scalar degree of freedom in addition to the two tensor modes.
This scalar mediates a new long-range force, i.e., a fifth force.
Since any deviation from general relativity is strongly constrained
in the solar system~\cite{Bertotti:2003rm,Will:2014kxa},
scalar-tensor theories must possess a mechanism for screening
the fifth force in the vicinity of matter sources such as in the solar system.

Several types of screening mechanisms have been known so far.
The first one relies on the potential term of the scalar degree of freedom.
The shape of the potential is designed so that the scalar becomes
effectively massive in a high density region.
This class of models include
chameleon~\cite{Khoury:2003rn}, symmetron~\cite{Hinterbichler:2010es},
and dilaton~\cite{Brax:2010gi} mechanisms.
The second one relies on nonlinear derivative interactions of the scalar field,
by which the kinetic term of the scalar becomes effectively large
and hence it is effectively weakly coupled to matter
near the source where its gradient is large.
This class can be divided into two subclasses
depending on whether first or second derivatives of the scalar field
play a crucial role. The former includes models
of~\cite{Burrage:2014uwa,Brax:2012jr,Babichev:2009ee}
and the latter includes the Galileons~\cite{Nicolis:2008in,Deffayet:2009wt,Deffayet:2009mn}.
The screening mechanism in
this last class of models is called the Vainshtein mechanism~\cite{Vainshtein:1972sx},
and has been studied extensively (see Ref.~\cite{Babichev:2013usa} for a review).
The Vainshtein mechanism has been
investigated~\cite{Kimura:2011dc,Narikawa:2013pjr,Koyama:2013paa,Kase:2013uja}
even in the context of the most general scalar-tensor theory
with second-order field equations~\cite{Horndeski:1974wa}
because it can be obtained by generalizing the Galileons~\cite{Deffayet:2011gz,Kobayashi:2011nu}
and the mechanism can thus be implemented naturally.
See Refs.~\cite{Kobayashi:2014ida,Crisostomi:2017lbg,Langlois:2017dyl,Dima:2017pwp}
for the Vainshtein mechanism (and its partial breaking)
in more general scalar-tensor theories which have been developed recently.


Previous works mostly focused on
the Vainshtein mechanism around spherical distributions of matter,
as a star can be well approximated by a sphere.
The authors of~\cite{Bloomfield:2014zfa}
have investigated analytically the systems with cylindrical
and planar symmetries, and found that
screening is weaker in the cylindrically symmetric case
and does not occur in the system with planar symmetry.
This implies that Vainshtein screening might be sensitive to
the shape of the matter distribution.
It is, however, difficult in general to study the shape dependence
of the Vainshtein mechanism because one has to treat derivative nonlinearities
in less symmetric systems.
In Ref.~\cite{Hiramatsu:2012xj},
a two-body system was investigated numerically and it was shown that
the equivalence principle can be violated apparently in such systems.
Approximate solutions for slowly rotating stars in the cubic Galileon theory
were obtained in Ref.~\cite{Chagoya:2014fza}.
As for a dynamical aspect of the Vainshtein mechanism,
the emission of scalar modes from a binary system
was evaluated in Ref.~\cite{deRham:2012fg}.
Very recently the shape dependence of screening
in the chameleon theory was addressed numerically in Ref.~\cite{Burrage:2017shh}.

In this paper, we consider a disk with a hole at its center
as a source and solve the Galileon field equation fully numerically
in order to address the consequence of nonlinear derivative interactions
in a less symmetric system.
We only study the cubic Galileons for simplicity.
A similar system in a different theory of modified gravity
has been considered in Refs.~\cite{Davis:2014tea,Davis:2016avf},
where scalar field profiles around a black hole accretion disk
have been investigated in the context of the chameleon theory.

The outline of this paper is as follows.
In the next section, we summarize briefly the cubic Galileon theory and
describe our numerical setup.
We then present our main results in Sec.~III.
Finally, we draw our conclusions in Sec.~IV.
In the main text we only consider the Galileon field living in a
flat background. To see how our result depends on the background curvature,
in Appendix~\ref{appe1} we give a numerical result in the
fixed Schwarzschild background.
Details of the numerical scheme are given in Appendix~\ref{appe2}.

\section{Basic equations}
\subsection{The cubic Galileon}

We consider the cubic Galileon
theory~\cite{Nicolis:2008in, Deffayet:2009wt}
as an example of the model endowed with the Vainshtein mechanism.
In the Einstein frame, the cubic Galileon $\phi$ and its coupling to matter
are described by the action
\begin{align}
 S=\int d^4x \left[-\frac{1}{2}(\partial\phi)^2-\frac{c_3}{M^3}(\partial\phi)^2\Box\phi
 +\frac{\beta}{M_{\rm Pl}}\phi T_\mu^{\;\mu}\right],\label{eq:action}
\end{align}
where $c_3$ and $\beta$ are dimensionless parameters,
$M_{\rm Pl}$ is the Planck mass, $M$ is another mass scale,
and $T_\mu^{\;\mu}$ is the trace of the matter energy-momentum tensor.
We assume that matter is non-relativistic, so that $T_\mu^{\;\mu}\simeq -\rho$.
Varying the action with respect to $\phi$, we obtain the field equation,
\begin{align}
 \Delta\phi +\frac{c_3}{M^3}\left[
 (\Delta\phi)^2-\nabla_i\nabla_j\phi\nabla^i\nabla^j\phi
 \right]=\frac{\beta}{M_{\rm Pl}}\rho,\label{eq:field}
\end{align}
where we assumed that $\phi$ is static.
For a given configuration of matter, one can integrate Eq.~(\ref{eq:field})
to obtain the profile of $\phi$.

The coupling between matter and the metric fluctuations around the Minkowski spacetime,
$h_{\mu\nu}$,
is expressed as $(1/2)h_{\mu\nu}T^{\mu\nu}$. This implies that the Jordan frame metric
is given by $h^{\rm J}_{\mu\nu}=h_{\mu\nu}+(2\beta/M_{\rm Pl})\phi\eta_{\mu\nu}$,
and thus a test particle of mass $m$ feels the fifth force
\begin{align}
 \Vec{F}_\phi=-\frac{\beta}{M_{\rm Pl}}m\Vec{\nabla}\phi,
\end{align}
in addition to the usual gravitational force $\Vec{F}_{\rm grav}=(m/2)\Vec{\nabla}h_{00}$.

Let us consider the profile of $\phi$ around a spherical matter configuration.
In this case, using the spherical coordinates it is easy to get
\begin{align}
 \Vec{\nabla}\phi = \frac{M^3}{4c_3}r\left(
 -1+\sqrt{1+\frac{2c_3\beta}{\pi M^3M_{\rm Pl}}\frac{{\cal M}}{r^3}}
 \right)\Vec{e}_r,
\end{align}
where $\Vec{e}_r$ is the radial unit vector and
${\cal M}$ is the mass of the spherical body.
For $r\gg r_{\rm V}:=(2c_3\beta{\cal M}/\pi M^3 M_{\rm Pl})^{1/3}$,
we have $|\Vec{F}_\phi|={\cal O}(\beta^2|\Vec{F}_{\rm grav}|)$,
implying that the fifth force is as large as the usual gravitational force if $\beta={\cal O}(1)$.
However, for $r\ll r_{\rm V}$ we find that
$|\Vec{F}_\phi|=(r/r_{\rm V})^{3/2}{\cal O}(\beta^2|\Vec{F}_{\rm grav}|)\ll {\cal O}(|\Vec{F}_{\rm grav}|)$, and thus the fifth force is screened in the vicinity of the body.
This is the Vainshtein mechanism.
For this to happen the non-linear term in Eq.~(\ref{eq:field}) plays a crucial role.
In order to pass laboratory and solar-system tests of gravity,
$M^3/c_3$ must be sufficiently small so that $r_{\rm V}$ is sufficiently large.

So far, successful Vainshtein screening has been confirmed
mainly for spherically symmetric configurations.
The Vainshtein mechanism in the systems with planar and cylindrical symmetry
has been investigated in Ref.~\cite{Bloomfield:2014zfa}, and it was found that
the screening of the fifth forth is sensitive to the shape of the matter distribution.
Only in such highly symmetric cases the non-linear equation~(\ref{eq:field}) can be
integrated analytically, and one has to employ numerical methods in general cases.
In Ref.~\cite{Hiramatsu:2012xj} the Galileon field equation is integrated numerically for
a two-body system.
In the present paper, we examine
the profile of the Galileon field around
a matter distribution that has not been investigated previously, i.e.,
a disk with a hole.

\subsection{Numerical setup}

\begin{figure}[tbp]
 \begin{center}
  \includegraphics[keepaspectratio=true,width=80mm]{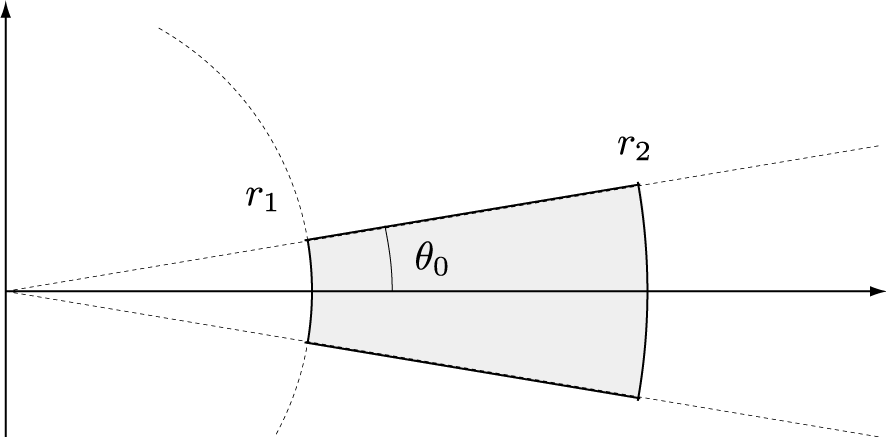}
 \end{center}
 \caption{A disk object with a hole in spherical coordinates.}%
 \label{fig:fig1.eps}
\end{figure}

Specifically, we model the system by the following uniform density profile
\begin{align}
 \rho(r,\theta) & =\rho_{0} U(r-r_{1})U(r_{2}-r)U(\theta_{0}-\theta)U(\theta_0+\theta),
 \\
 \rho_0         & ={\rm const},
\end{align}
where $U$ is the Heaviside function, with $r_1$, $r_2$, and $\theta_0$
being the inner radius, the outer radius, and the
(half of the)
opening angle of the disk,
respectively (Fig.~\ref{fig:fig1.eps}).
Note that here we are using the
spherical coordinates whose definition is slightly different from
the usual one,
$x=r\cos\theta\cos\varphi$, $y=r\cos\theta\sin\varphi$, $z=r\sin\theta$,
with $-\pi/2\le \theta\le \pi/2$.

To implement numerical integration, we introduce
the following dimensionless quantities:
\begin{align}
 \bar{\phi}:=\frac{\phi}{M^{3}r_{0}^{2}}, \quad \bar{r}:=\frac{r}{r_{0}},
 \quad \mu:= \frac{\beta \rho_{0}}{M^{3}M_{{\rm Pl}}},
\end{align}
where $r_0$ is some arbitrary length scale and
$\mu$ is the parameter that corresponds to the coupling between matter and the Galileon
for fixed $\rho_0$.
At a sufficiently large distance from the disk object,
it can be regarded as a point particle and hence
we have
$\bar\phi\sim \mu /\bar r^2$. Therefore, it can be said that $\mu$ controls the nonlinearity
of the scalar field.
We rewrite Eq.~(\ref{eq:field}) in terms of the above variables
assuming that $\phi$ is axisymmetric.

The boundary conditions we impose are given by
\begin{align}
   & \left.\frac{\henbi \bar{\phi}}{\henbi \bar{r}}\right|_{\bar{r}=0}=0,\label{eq:regularity}
 \\
   & \left.\frac{\henbi \bar{\phi}}{\henbi \theta}\right|_{\theta=0}=\left.\frac{\henbi \bar{\phi}}{\henbi \theta}\right|_{\theta=\pi/2}=0,\label{eq:simmetry}
 \\
   & \;\bar{\phi}(\bar{r}_{{\rm max}},\theta)=0,\label{eq:galileon}
\end{align}
where $\bar r_{\rm max}:=r_{\rm max}/r_0$
corresponds to the boundary of the computational domain.
The boundary condition~(\ref{eq:regularity}) amounts to the regularity at the center,
while the condition~(\ref{eq:simmetry}) reflects the symmetry of the system.
Since the field equation is invariant under the constant shift of the scalar field,
$\phi\to \phi + c$, we may impose the boundary condition~(\ref{eq:galileon})
without loss of generality.

One may naively expect that derivative nonlinearity of
the Galileon field is large for $r\lesssim (c_3 \beta \rho_0 V/M^2M_{\rm Pl})^{1/3}$,
where $V$ is the size of a massive object. If we roughly take $r_1\sim r_2$,
we can estimate $V$ as $V\sim r_2^3\theta_0$. Thus, in terms of
the dimensionless variables, we see that the nonlinear effect is large for
$\bar{r}\lesssim (c_3\mu \theta_0)^{1/3}\bar{r}_2$.

\section{Numerical Results}

We now present our numerical solutions to Eq.~(\ref{eq:field}).
We fix $\bar{r}_2$ and $\bar{r}_{\rm max}$ as
$\bar{r}_2= 30$ and $\bar{r}_{\rm max}=80$, respectively,
and performed numerical calculations for different values of
$r_1$, $\theta_0$, and $\mu$.
The number of data points is 200 in the $r$ direction
and 100 in the $\theta$ direction.
The details of the numerical computation are described in the Appendix~\ref{appe2}.

In Fig.~\ref{fig:vector1}
we show a vector plot of the dimensionless
force field $-(M^3r_0)^{-1}\Vec{\nabla}\phi$ for $c_3=1$,
$\bar{r}_1=8$, $\theta_0=0.05$, and $\mu=36.8$.
In order to clarify the effect of the nonlinear terms in Eq.~(\ref{eq:field}),
we also calculated the force field
with the same parameters, but with $c_3=0$. The result is also presented in
Fig.~\ref{fig:vector1} for comparison.
It can be seen that in the $c_3=1$ case the fifth force is suppressed
compared to the $c_3=0$ case
in almost every region, as expected.
This is clear in particular for $\bar{r}\gtrsim 20$ around the disk.
However, surprisingly enough,
the nonlinear effect {\em enhances}, rather than suppresses,
the fifth force in the vicinity of the hole.

To quantify this anti-screening effect,
we introduce the following scalar quantity,
\begin{align}
 \mathcal{R}=\frac{|\Vec{\nabla} \phi|_{c_{3}=1}}{|\Vec{\nabla} \phi|_{c_{3}=0}}.
 \label{eq:def_R}
\end{align}
We may say that screening is successful if ${\cal R}<1$.
Figure~\ref{fig:main_res} shows ${\cal R}$ for the above case,
which clearly indicates that the fifth force is enhanced near the hole.

\begin{figure}[!htbp]
 \centering
 \includegraphics[width=8cm]{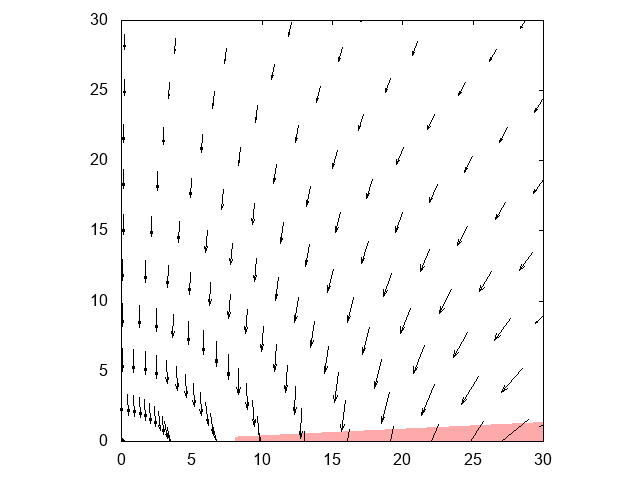}
 \includegraphics[width=8cm]{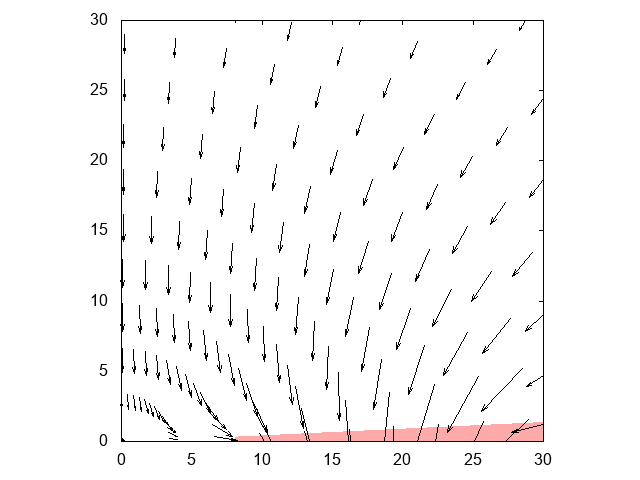}
 \caption{Dimensionless force fields for
 $\bar{r}_1=8$, $\theta_0=0.05$, and $\mu=36.8$,
 with $c_3=1$ (left) and $c_3=0$ (right). The thin black region
 represents the disk.
 }
 \label{fig:vector1}
\end{figure}

\begin{figure}[!h]
 \centering
 \includegraphics[width=9.5cm]{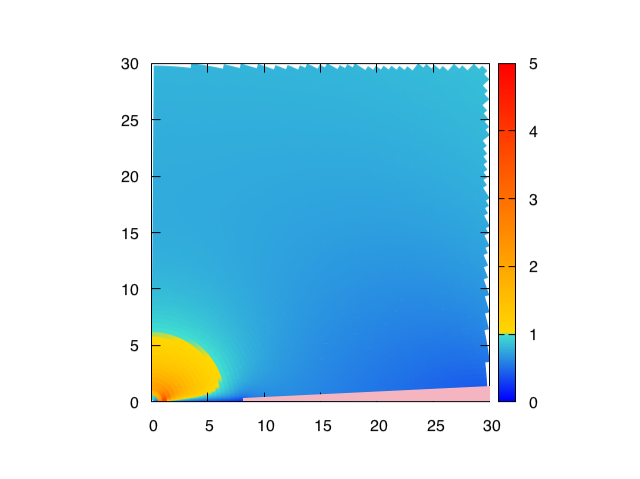}
 \includegraphics[width=8cm]{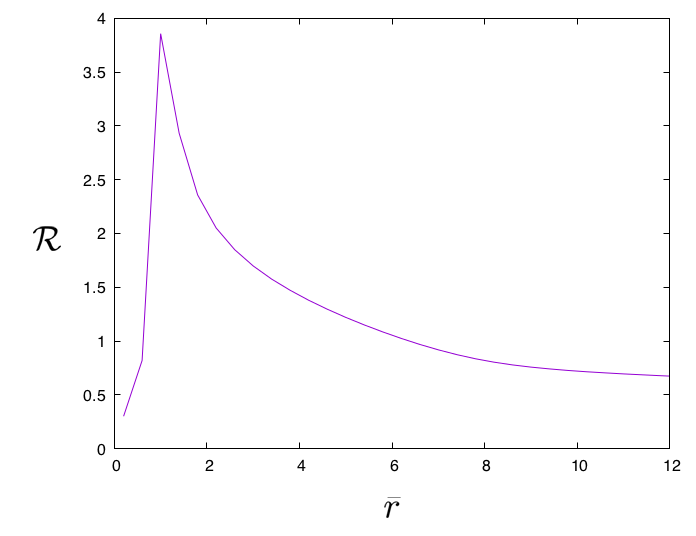}
 \caption{2D plot of the degree of (anti-)screening
 ${\cal R}$ for the case shown in Fig.~\ref{fig:vector1}
 (left) and ${\cal R}$ along $\theta = 2\pi/5$ as a function of $\bar{r}$ (right).
 }
 \label{fig:main_res}
\end{figure}

\begin{figure}[h]
 \centering
 \includegraphics[width=8.5cm]{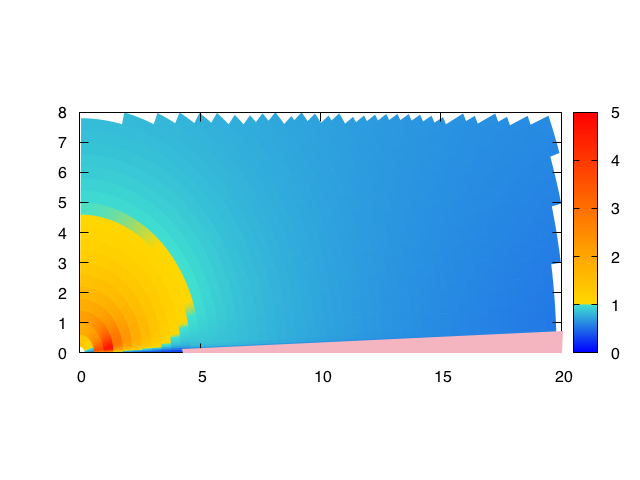}
 \includegraphics[width=8.5cm]{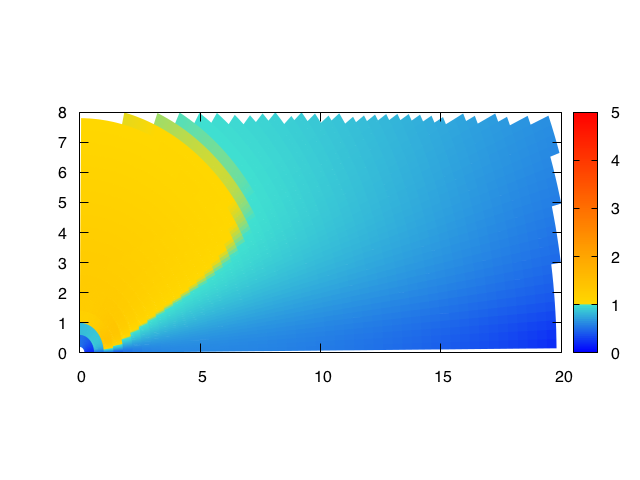}
 \includegraphics[width=9cm]{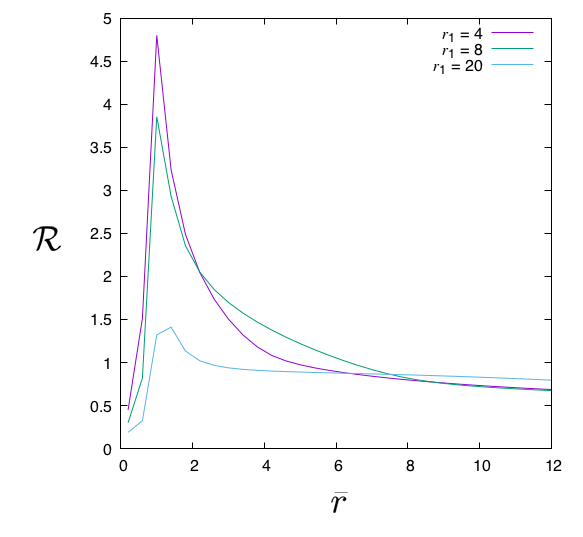}
 \caption{$\mathcal{R}$ for $r_{1}/r_0=4$ (top left) and $r_{1}/r_0=20$ (top right).
 The other parameters are the same as in the previous plots.
 ${\cal R}$ along $\theta = 2\pi/5$ as a function of $\bar{r}$
is also shown (bottom). }
 \label{rdepen}
\end{figure}

\begin{figure}[h]
 \centering
 \includegraphics[width=8.5cm]{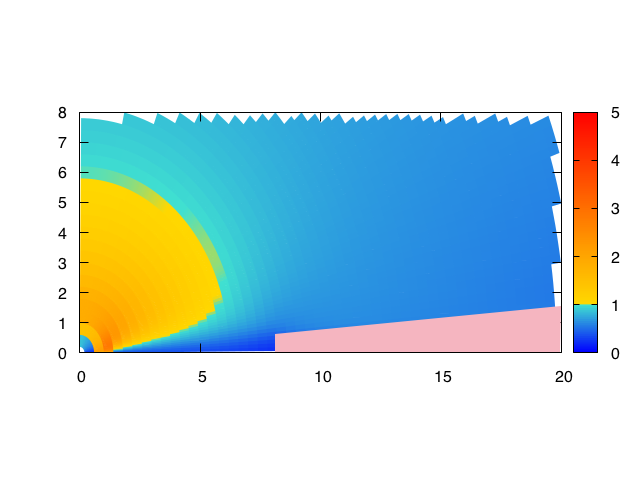}
 \includegraphics[width=8.5cm]{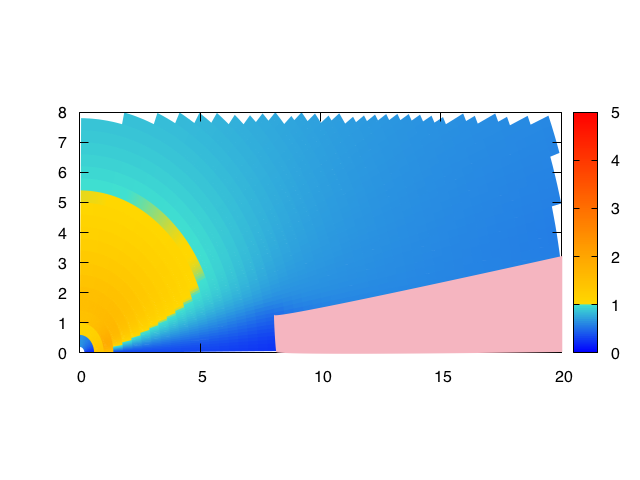}
 \includegraphics[width=9cm]{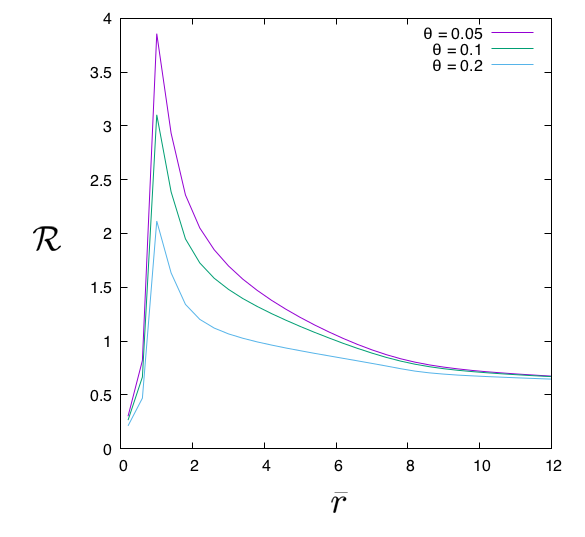}
 \caption{$\mathcal{R}$ for $\theta_{0}=0.1$ (top left)
 and $\theta_{0}=0.2$ (top right). ${\cal R}$ along
 $\theta = 2\pi/5$ as a function of $\bar{r}$
is also shown (bottom).}
 \label{thetadepen}
\end{figure}

\begin{figure}[h]
 \centering
 \includegraphics[width=8.5cm]{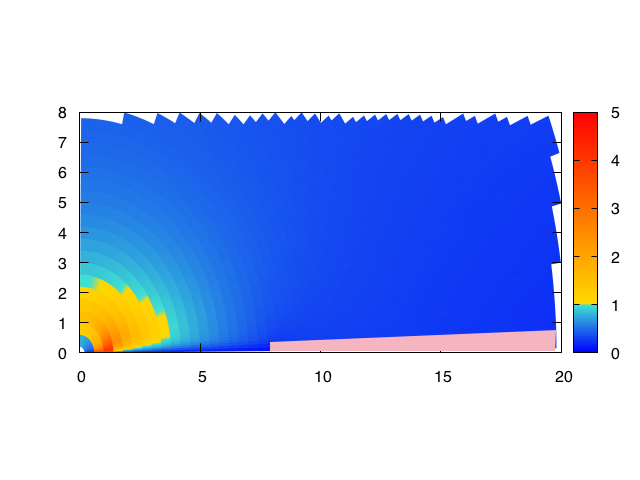}
 \includegraphics[width=8.5cm]{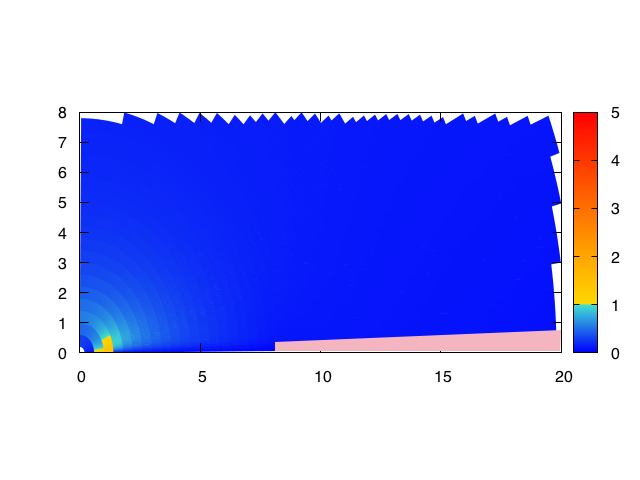}
 \includegraphics[width=9cm]{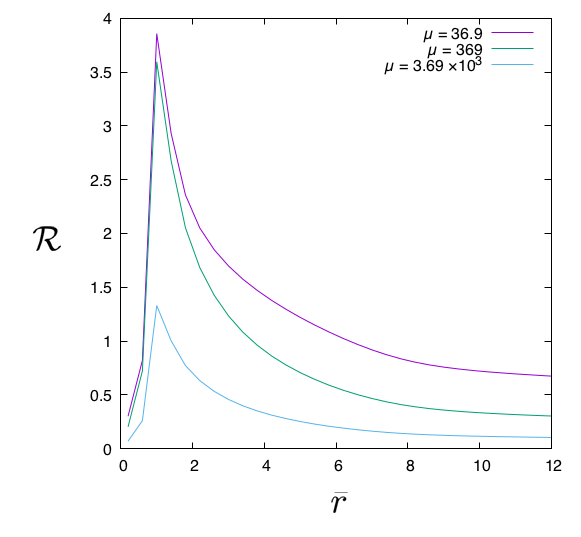}
 \caption{$\mathcal{R}$ for $\mu=369$ (left) and $\mu=3690$ (right).
 ${\cal R}$ along
 $\theta = 2\pi/5$ as a function of $\bar{r}$
is also shown (bottom).}
 \label{mudepen}
\end{figure}

To see how the enhancement of the fifth force depends on the parameters,
we provide
numerical results for different values of $r_1$, $\theta_0$ and $\mu$
in Figs.~\ref{rdepen}--\ref{mudepen}.
Figure~\ref{rdepen} shows ${\cal R}$ for different sizes of the hole,
$\bar{r}_1=4$ and $20$, with $c_3$, $r_1$, $\theta_0$ being fixed
to the previous values and $\mu$ being given such that the total mass of
the disk is unchanged from the previous case.
It is found that
the fifth force around the hole is stronger for a smaller hole size,
as is most clearly seen in the bottom panel of Fig.~\ref{rdepen}.
Figure~\ref{thetadepen} represents the dependence of
the enhancement effect on the thickness of the disk.
We see that for smaller $\theta_0$, i.e., for a thinner disk,
the fifth force around the hole is stronger.
Finally, we see from Fig.~\ref{mudepen} how increasing $\mu$
changes the result with other parameters fixed.
For larger $\mu$, the enhancement of the Galileon force is
less evident. This is because larger $\mu$ implies
that the disk is (effectively) more dense or more massive,
and thus the screening effect from the disk itself is more efficient.
To sum up, the anti-screening effect is
larger for a thinner, less massive disk with a smaller hole.

\section{Discussion}

In this paper, we have studied numerically the fifth force
around a disk with a hole at its center
in the cubic Galileon theory.
It is known that Vainshtein screening does not work
for infinite planar sources~\cite{Bloomfield:2014zfa}.
Since our source is thin but finite, we have seen that
screening still occurs
in almost every region around the disk.
However, we have found that
the hole at the center causes an unexpected consequence:
the Galileon force is not suppressed but
enhanced in the vicinity of the hole,
namely, anti-screening operates.
Anti-screening we have seen in this paper occurs in the region where
nonlinearity of the Galileon field is dominant and
the configuration of matter is less symmetric.
Due to this complexity, so far we have not arrived at analytic understanding
of our result.

Some of the parameters we have used in our numerical calculations
might not be realistic. In particular, we have seen that we need $\mu \lesssim 10^3$
in order for the force to be enhanced.
For larger $\mu$, the effect of anti-screening is washed away by
the screening effect from the disk in the present setup.
If the Galileon field is responsible for the current cosmic acceleration,
one would expect $M^3\sim M_{\rm Pl} H_0^2 \sim \bar\rho/M_{\rm Pl}$,
where $H_0$ is the present Hubble parameter and $\bar\rho$
is the average energy density of the Universe.
The energy density of our disk is thus given by $\rho_0\sim \mu\bar\rho$,
assuming that $\beta={\cal O}(1)$.
Our numerical calculations correspond to such
very low density matter distribution.
Therefore, at this stage it is difficult to derive
direct implications of our results
for astrophysics and experiments.
Nevertheless,
we believe that it is interesting to further explore
how the (anti-)screening mechanism operates for nontrivial configurations
of matter and the present work provides a first step toward
understanding this complicated problem.

\acknowledgments
We thank Christos Charmousis,
A. Emir G\"umr\"uk\c c\"uo\u glu, Tomohiro Harada, and Kazuya Koyama for useful comments.
H.O. also thanks Kazuya Koyama for his kind hospitality
at University of Portsmouth
where part of this work was done.
This work was supported in part by
the Rikkyo University Special Fund for Research (H.O.),
JSPS Overseas Challenge Program for Young Researchers (H.O.),
the JSPS Grants-in-Aid for Scientific Research
 Nos.~16K17695 (T.H.),
16H01102, and 16K17707 (T.K),
MEXT-Supported Program for the Strategic Research Foundation at Private Universities,
2014-2017 (S1411024) (T.H. and T.K.),
and MEXT KAKENHI Grant Nos.~15H05888 (T.K.) and~17H06359 (T.K.).

\appendix

\section{Scalar-field profile in Schwarzschild geometry}\label{appe1}

In the main text we have solved the Galileon field equation
in the flat background. In order to see the scalar-field profile
in the curved background, let us consider the covariant version of
Eq.~(\ref{eq:field}) in a {\em fixed} background:
\begin{align}
\Box\phi +\frac{c_3}{M^3}\left[
(\Box\phi)^2-\nabla_\mu\nabla_\nu\phi\nabla^\mu\nabla^\nu\phi
-R^{\mu\nu}\nabla_\mu\phi\nabla_\nu\phi
\right] = \frac{\beta}{M_{\rm Pl}}\rho,
\label{eq:curved}
\end{align}
where $\nabla_\mu$ is the covariant derivative with respect to
$g_{\mu\nu}$, $R_{\mu\nu}$ is the Ricci tensor, and we take
\begin{align}
g_{\mu\nu}dx^\mu dx^\nu=-\left(1-\frac{r_g}{r}\right)dt^2 +
\left(1-\frac{r_g}{r}\right)^{-1}dr^2 +r^2d \Omega_2^2,
\end{align}
with $r_g=r_0$.
We solve Eq.(\ref{eq:curved}) numerically for
the matter configuration with $\bar{r}_1=8$, $\theta_0=0.05$,
and $\mu=36.8$.
The resultant profile of $\phi$ should be compared with that in the flat background
in Fig.~\ref{fig:main_res}. It can be seen that the profiles
are not so different. We thus conclude
that the effect of the background curvature does not
change the result of anti-screening.

\begin{figure}[htbp]
 \centering
 \includegraphics[width=9.5cm]{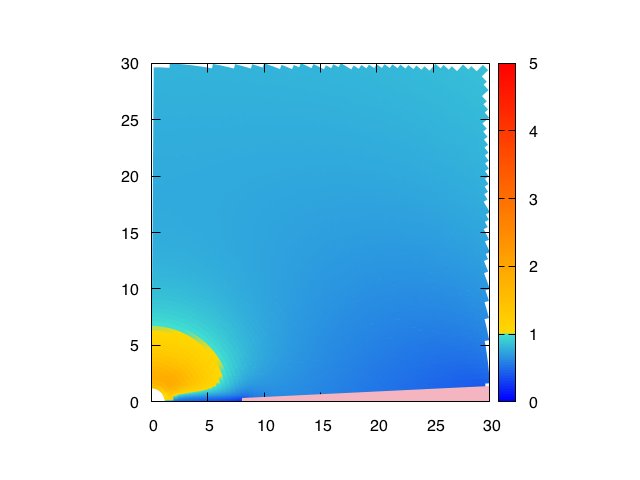}
 \caption{$\mathcal{R}$ in the Schwarzschild background.
 This should be compared with the left panel of Fig.~\ref{fig:main_res}.
 }
 \label{fig:BH}
\end{figure}

\section{Numerical scheme and convergence of results}\label{appe2}

Throughout this paper, we employed the numerical scheme developed in
Ref.~\cite{Hiramatsu:2012xj} to solve the field equation
(\ref{eq:field}).
Basically, we regard the non-linear terms
of Eq.~(\ref{eq:field}), the terms proportional to $c_3$,
as the extra source term such that
\begin{align}
 \triangle\phi = \frac{\beta}{M_{\rm Pl}}\rho - \frac{c_3}{M^3}N[\phi].
\end{align}
At the first step, we solve the linear equation with setting $N[\phi]=0$ and
obtain the solution $\phi_*$. Then we update $\phi$ in the following
manner:
\begin{align}
 \phi_{\rm new}(r,\theta) = (1-\omega)\phi_{\rm old}(r,\theta) + \omega\phi_*(r,\theta),
\end{align}
with a mixing parameter $\omega=\mathcal{O}(0.01)$.
At the next step, evaluating $N[\phi_{\rm new}]$, we solve the field
equation again, and $\phi$ is futher updated. This iteration procedure
is terminated when the update of $\phi$ is well suppressed, namely,
\begin{align}
 \frac{||\phi_{\rm new}-\phi_{\rm old}||}{||\phi_{\rm new}||} < 10^{-8},
\end{align}
where the norm $||\phi||$ is defined as $||\phi||:=\sqrt{\sum_{ij}\phi(r_i,\theta_j)^2}$.
Note that, unless the parameter $\omega$ is small, this iteration scheme
does not work since the non-linear term $N[\phi]$ induces quite a large
change of the field configuration. For details, see
Ref.~\cite{Hiramatsu:2012xj}.

The field equation solved in this paper, given in Eq.~(\ref{eq:field}),
is highly non-linear, and thus it should be confirmed whether our
numerical results are reliable in the sense that they are well
converged with the iteration scheme mentioned above.
To see this, we solve the field equation with changing the number of grid points in the coordinate of
$(r,\theta)$, $N_r$ and $N_\theta$, and the position of the boundary in
the radial direction, $\bar{r}_{\rm max}$. The fiducial values of $N_r$, $N_\theta$ and
$\bar{r}_{\rm max}$ in this paper are $N_r=200, N_\theta=100$ and
$\bar{r}_{\rm max}=80$. Note that we do not focus on the other numerical
parameters since they control solely the convergence
speed and the precision, and thus do not affect the final results.

\begin{figure}[htbp]
 \centering
 \includegraphics[width=5.4cm]{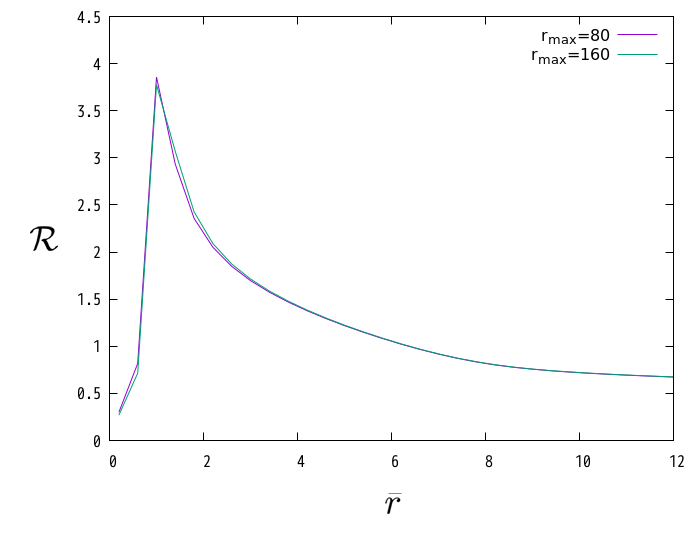}
 \includegraphics[width=5cm]{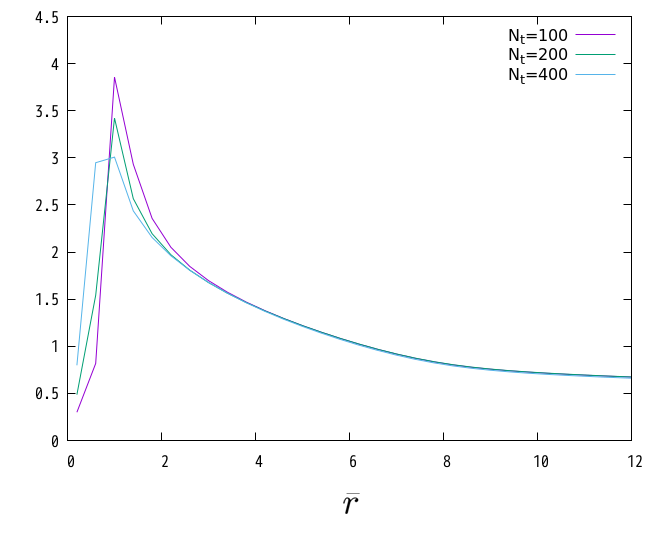}
 \includegraphics[width=5cm]{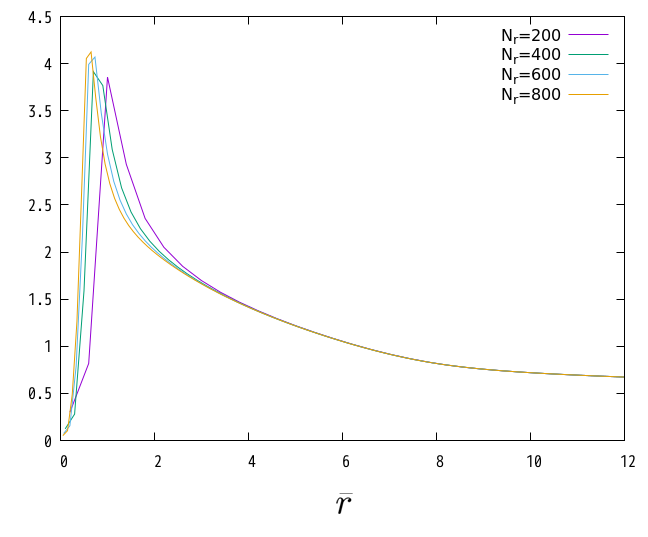}
 \caption{The convergence check of numerical results; the dependence on
  $\bar{r}_{\rm max}$ (left), on $N_r$ (middle) and on $N_\theta$ (right).}
 \label{fig:convcheck}
\end{figure}

Figure \ref{fig:convcheck} shows $\mathcal{R}$ evaluated at
$\theta = 2\pi/5$. From the left panel, we find that our result
is insensitive to the size of the computational box, which means that
artificial effects from the boundary at $r=r_{\rm max}$ do not affect
the feature at all. The remaining panels show the dependences of
$\mathcal{R}$ on the
number of grids, $N_r$ (center panel) and $N_\theta$ (right panel.)
While the detailed structure of the peak around $\bar{r}\sim 1$ is
sensitive to the spatial resolution, the fact that $\mathcal{R}$ can be
larger than the unity at $\bar{r}\lesssim 8$ is confirmed to be robust.

\begin{thebibliography}{99}

 \bibitem{Perlmutter:1998np}
 S.~Perlmutter {\it et al.} [Supernova Cosmology Project Collaboration],
 ``Measurements of Omega and Lambda from 42 high redshift supernovae,''
 Astrophys.\ J.\  {\bf 517}, 565 (1999)
 [astro-ph/9812133].

 \bibitem{Riess:1998cb}
 A.~G.~Riess {\it et al.} [Supernova Search Team],
 ``Observational evidence from supernovae for an accelerating universe and a cosmological constant,''
 Astron.\ J.\  {\bf 116}, 1009 (1998)
 [astro-ph/9805201].

 \bibitem{Spergel:2003cb}
 D.~N.~Spergel {\it et al.} [WMAP Collaboration],
 ``First year Wilkinson Microwave Anisotropy Probe (WMAP) observations: Determination of cosmological parameters,''
 Astrophys.\ J.\ Suppl.\  {\bf 148}, 175 (2003)
 [astro-ph/0302209].

 \bibitem{Eisenstein:2005su}
 D.~J.~Eisenstein {\it et al.} [SDSS Collaboration],
 ``Detection of the Baryon Acoustic Peak in the Large-Scale Correlation Function of SDSS Luminous Red Galaxies,''
 Astrophys.\ J.\  {\bf 633}, 560 (2005)
 [astro-ph/0501171].

 \bibitem{Abazjian:2008wr}
 K.~N.~Abazajian {\it et al.} [SDSS Collaboration],
 ``The Seventh Data Release of the Sloan Digital Sky Survey,''
 Astrophys.\ J.\ Suppl.\  {\bf 182}, 543 (2009)
 [arXiv:0812.0649 [astro-ph]].

 \bibitem{Ade:2013zuv}
 P.~A.~R.~Ade {\it et al.} [Planck Collaboration],
 ``Planck 2013 results. XVI. Cosmological parameters,''
 Astron.\ Astrophys.\  {\bf 571}, A16 (2014)
 [arXiv:1303.5076 [astro-ph.CO]].



 \bibitem{Copeland:2006wr}
 See, {\em e.g.}, E.~J.~Copeland, M.~Sami and S.~Tsujikawa,
 ``Dynamics of dark energy,''
 Int.\ J.\ Mod.\ Phys.\ D {\bf 15}, 1753 (2006),
 [hep-th/0603057].

 \bibitem{clifton}
 See, {\em e.g.}, T.~Clifton,~P.~G.~Ferreira,~A.~Padilla and C.~Skordis,
 ``Modified gravity and cosmology,''
 Phys.\ Rep.,\ {\bf 513},\ 1  (2012), [arXiv:1106.2476 [astro-ph.CO]].

   \bibitem{Will:2014kxa}
  C.~M.~Will,
  ``The Confrontation between General Relativity and Experiment,''
  Living Rev.\ Rel.\  {\bf 17}, 4 (2014)
  [arXiv:1403.7377 [gr-qc]].

 \bibitem{Bertotti:2003rm}
 B.~Bertotti, L.~Iess and P.~Tortora,
 ``A test of general relativity using radio links with the Cassini spacecraft,''
 Nature {\bf 425}, 374 (2003).

 \bibitem{Khoury:2003rn}
 J.~Khoury and A.~Weltman,
 ``Chameleon cosmology,''
 Phys.\ Rev.\ D {\bf 69}, 044026 (2004)
 [astro-ph/0309411].

 \bibitem{Hinterbichler:2010es}
 K.~Hinterbichler and J.~Khoury,
 ``Symmetron Fields: Screening Long-Range Forces Through Local Symmetry Restoration,''
 Phys.\ Rev.\ Lett.\  {\bf 104}, 231301 (2010)
 [arXiv:1001.4525 [hep-th]].

 \bibitem{Brax:2010gi}
 P.~Brax, C.~van de Bruck, A.~C.~Davis and D.~Shaw,
 ``The Dilaton and Modified Gravity,''
 Phys.\ Rev.\ D {\bf 82}, 063519 (2010)
 [arXiv:1005.3735 [astro-ph.CO]].
 \bibitem{Burrage:2014uwa}
 C.~Burrage and J.~Khoury,
 ``Screening of scalar fields in Dirac-Born-Infeld theory,''
 Phys.\ Rev.\ D {\bf 90}, no. 2, 024001 (2014)
 [arXiv:1403.6120 [hep-th]].


 \bibitem{Babichev:2009ee}
 E.~Babichev, C.~Deffayet and R.~Ziour,
 ``k-Mouflage gravity,''
 Int.\ J.\ Mod.\ Phys.\ D {\bf 18}, 2147 (2009)
 [arXiv:0905.2943 [hep-th]].

 \bibitem{Brax:2012jr}
 P.~Brax, C.~Burrage and A.~C.~Davis,
 ``Screening fifth forces in k-essence and DBI models,''
 JCAP {\bf 1301}, 020 (2013)
 [arXiv:1209.1293 [hep-th]].

 \bibitem{Nicolis:2008in}
 A.~Nicolis, R.~Rattazzi and E.~Trincherini,
 ``The Galileon as a local modification of gravity,''
 Phys.\ Rev.\ D {\bf 79}, 064036 (2009)
 [arXiv:0811.2197 [hep-th]].

 \bibitem{Deffayet:2009wt}
 C.~Deffayet, G.~Esposito-Farese and A.~Vikman,
 ``Covariant Galileon,''
 Phys.\ Rev.\ D {\bf 79}, 084003 (2009)
 [arXiv:0901.1314 [hep-th]].

 \bibitem{Deffayet:2009mn}
 C.~Deffayet, S.~Deser and G.~Esposito-Farese,
 ``Generalized Galileons: All scalar models whose curved background extensions maintain
 second-order field equations and stress-tensors,''
 Phys.\ Rev.\ D {\bf 80}, 064015 (2009)
 [arXiv:0906.1967 [gr-qc]].


 \bibitem{Vainshtein:1972sx}
 A.~I.~Vainshtein,
 ``To the problem of nonvanishing gravitation mass,''
 Phys.\ Lett.\  {\bf 39B}, 393 (1972).

         \bibitem{Babichev:2013usa}
           E.~Babichev and C.~Deffayet,
           ``An introduction to the Vainshtein mechanism,''
           Class.\ Quant.\ Grav.\  {\bf 30}, 184001 (2013)
           [arXiv:1304.7240 [gr-qc]].

 \bibitem{Kimura:2011dc}
 R.~Kimura, T.~Kobayashi and K.~Yamamoto,
 ``Vainshtein screening in a cosmological background in the most general second-order scalar-tensor theory,''
 Phys.\ Rev.\ D {\bf 85}, 024023 (2012)
 [arXiv:1111.6749 [astro-ph.CO]].

 \bibitem{Narikawa:2013pjr}
 T.~Narikawa, T.~Kobayashi, D.~Yamauchi and R.~Saito,
 ``Testing general scalar-tensor gravity and massive gravity with cluster lensing,''
 Phys.\ Rev.\ D {\bf 87}, 124006 (2013)
 [arXiv:1302.2311 [astro-ph.CO]].


 \bibitem{Koyama:2013paa}
   K.~Koyama, G.~Niz and G.~Tasinato,
   ``Effective theory for the Vainshtein mechanism from the Horndeski action,''
   Phys.\ Rev.\ D {\bf 88}, 021502 (2013)
   [arXiv:1305.0279 [hep-th]].

 \bibitem{Kase:2013uja}
 R.~Kase and S.~Tsujikawa,
 ``Screening the fifth force in the Horndeski's most general scalar-tensor theories,''
 JCAP {\bf 1308}, 054 (2013)
 [arXiv:1306.6401 [gr-qc]].

 \bibitem{Horndeski:1974wa}
   G.~W.~Horndeski,
   ``Second-order scalar-tensor field equations in a four-dimensional space,''
   Int.\ J.\ Theor.\ Phys.\  {\bf 10}, 363 (1974).
 \bibitem{Deffayet:2011gz}
 C.~Deffayet, X.~Gao, D.~A.~Steer and G.~Zahariade,
 ``From k-essence to generalised Galileons,''
 Phys.\ Rev.\ D {\bf 84}, 064039 (2011)
 [arXiv:1103.3260 [hep-th]].

 \bibitem{Kobayashi:2011nu}
   T.~Kobayashi, M.~Yamaguchi and J.~Yokoyama,
   ``Generalized G-inflation: Inflation with the most general second-order field equations,''
   Prog.\ Theor.\ Phys.\  {\bf 126}, 511 (2011)
   [arXiv:1105.5723 [hep-th]].

     \bibitem{Kobayashi:2014ida}
       T.~Kobayashi, Y.~Watanabe and D.~Yamauchi,
       ``Breaking of Vainshtein screening in scalar-tensor theories beyond Horndeski,''
       Phys.\ Rev.\ D {\bf 91}, no. 6, 064013 (2015)
       [arXiv:1411.4130 [gr-qc]].

     \bibitem{Crisostomi:2017lbg}
       M.~Crisostomi and K.~Koyama,
       ``Vainshtein mechanism after GW170817,''
       Phys.\ Rev.\ D {\bf 97}, no. 2, 021301 (2018)
       [arXiv:1711.06661 [astro-ph.CO]].

       \bibitem{Langlois:2017dyl}
         D.~Langlois, R.~Saito, D.~Yamauchi and K.~Noui,
         ``Scalar-tensor theories and modified gravity in the wake of GW170817,''
         arXiv:1711.07403 [gr-qc].

       \bibitem{Dima:2017pwp}
         A.~Dima and F.~Vernizzi,
         ``Vainshtein Screening in Scalar-Tensor Theories before and after GW170817: Constraints on Theories beyond Horndeski,''
         arXiv:1712.04731 [gr-qc].



 \bibitem{Bloomfield:2014zfa}
 J.~K.~Bloomfield, C.~Burrage and A.~C.~Davis,
 ``Shape dependence of Vainshtein screening,''
 Phys.\ Rev.\ D {\bf 91}, no. 8, 083510 (2015)
 [arXiv:1408.4759 [gr-qc]].

 \bibitem{Hiramatsu:2012xj}
 T.~Hiramatsu, W.~Hu, K.~Koyama and F.~Schmidt,
 ``Equivalence Principle Violation in Vainshtein Screened Two-Body Systems,''
 Phys.\ Rev.\ D {\bf 87}, no. 6, 063525 (2013)
 [arXiv:1209.3364 [hep-th]].

 \bibitem{Chagoya:2014fza}
 J.~Chagoya, K.~Koyama, G.~Niz and G.~Tasinato,
 ``Galileons and strong gravity,''
 JCAP {\bf 1410}, no. 10, 055 (2014)
 [arXiv:1407.7744 [hep-th]].

 \bibitem{deRham:2012fg}
   C.~de Rham, A.~Matas and A.~J.~Tolley,
   ``Galileon Radiation from Binary Systems,''
   Phys.\ Rev.\ D {\bf 87}, no. 6, 064024 (2013)
   [arXiv:1212.5212 [hep-th]].

   \bibitem{Burrage:2017shh}
     C.~Burrage, E.~J.~Copeland, A.~Moss and J.~A.~Stevenson,
     ``The shape dependence of chameleon screening,''
     arXiv:1711.02065 [astro-ph.CO].

 \bibitem{Davis:2014tea}
 A.~C.~Davis, R.~Gregory, R.~Jha and J.~Muir,
 ``Astrophysical black holes in screened modified gravity,''
 JCAP {\bf 1408}, 033 (2014)
 [arXiv:1402.4737 [astro-ph.CO]].

 \bibitem{Davis:2016avf}
 A.~C.~Davis, R.~Gregory and R.~Jha,
 ``Black hole accretion discs and screened scalar hair,''
 JCAP {\bf 1610}, no. 10, 024 (2016)
 [arXiv:1607.08607 [gr-qc]].





\end{thebibliography}
\end{document}